\documentclass{aastex7}
\usepackage{siunitx}
\usepackage[aas,bfvec]{math-shortcuts}

\begin{document}

\title{A Generalized Template Matching Algorithm for Correcting Jitter Noise in Pulsar Timing}

\author[0000-0003-1082-2342]{Ross J. Jennings}
\email{ross.jennings@mail.wvu.edu}
\affiliation{Department of Physics and Astronomy, West Virginia University, P.O. Box 6315, Morgantown, WV 26506, USA}
\altaffiliation{NANOGrav Physics Frontiers Center Postdoctoral Fellow}

\author[0000-0002-4049-1882]{James M. Cordes}
\email{cordes@astro.cornell.edu}
\affiliation{Cornell Center for Astrophysics and Planetary Science and Department of Astronomy, Cornell University, Ithaca, NY 14853, USA}

\author[0000-0002-2878-1502]{Shami Chatterjee}
\email{shami@astro.cornell.edu}
\affiliation{Cornell Center for Astrophysics and Planetary Science and Department of Astronomy, Cornell University, Ithaca, NY 14853, USA}

\author[0000-0001-7697-7422]{Maura A. McLaughlin}
\email{maura.mclaughlin@mail.wvu.edu}
\affiliation{Department of Physics and Astronomy, West Virginia University, P.O. Box 6315, Morgantown, WV 26506, USA}

\begin{abstract}

Pulsar timing is a valuable source of high-precision astrophysical measurements which can be used to probe gravitational physics, including by detecting gravitational waves. An important factor limiting the precision of these measurements is pulse jitter; i.e., intrinsic, short-timescale variation in the amplitude and shape of pulses from a given pulsar. Because conventional pulse time-of-arrival (TOA) measurement relies on template matching, which assumes the average pulse shape is stable, such variation gives rise to jitter noise in TOA measurements. Here we introduce a generalization of the template matching technique, making use of principal component analysis, which can account for variations in pulse shape. We compare this technique to other proposals for mitigating jitter noise in pulsar timing, paying particular attention to the possibility of corrections absorbing other astrophysical signals of interest, and demonstrate its effectiveness using simulated data.

\end{abstract}

\section{Introduction}

The pulsar timing technique depends on the link between measured pulse times of arrival (TOAs) and the rotation phase of a pulsar. For millisecond pulsars (MSPs), which typically have rotational periods of a few milliseconds and spindown rates on the order of $10^{-20}$\,\si{s/s}, TOAs can often be measured with a precision of tens to hundreds of nanoseconds, and predicted accurately over time spans of years to decades. This high precision and accuracy enables a wide variety of scientific applications, including the detection of energy loss from gravitational-wave emission in binary systems \citep[e.g.,][]{tw82}, tests of the equivalence principle and other strong-field gravitational effects \citep[e.g.,][]{stairs03}, detections of small bodies orbiting pulsars \citep[e.g.,][]{wolszczan94}, measurement of neutron star masses via Shapiro delay \citep[e.g.,][]{cfr+20}, and studies of the structure of the ionized interstellar medium \citep[e.g.,][]{ne2001}. Ongoing efforts by pulsar timing arrays (PTAs; \citealt{dcl+16,hobbs13}), including the North American Nanohertz Observatory for Gravitational Waves (NANOGrav; \citealt{5yr-dataset,15yr-stochastic}), seek to use MSP timing to detect and characterize the sources of gravitational waves at nanohertz frequencies. Such sources are expected to include the stochastic background generated by supermassive black hole binaries, as well as signals from identifiable individual binary systems, and may also include signals produced by various forms of new physics~\citep{15yr-astrophysics,15yr-new-physics}.

\begin{figure}
\centering
\includegraphics[width=0.33\textwidth]{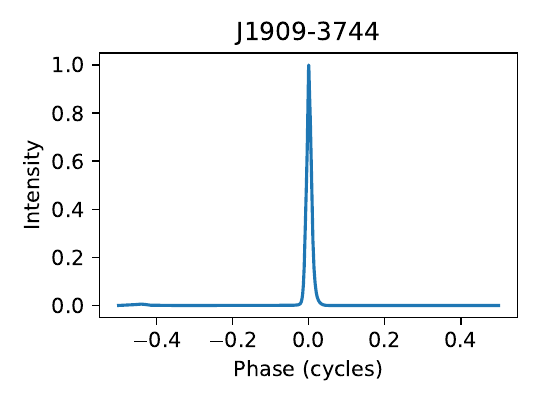}\hspace{-1em}
\includegraphics[width=0.33\textwidth]{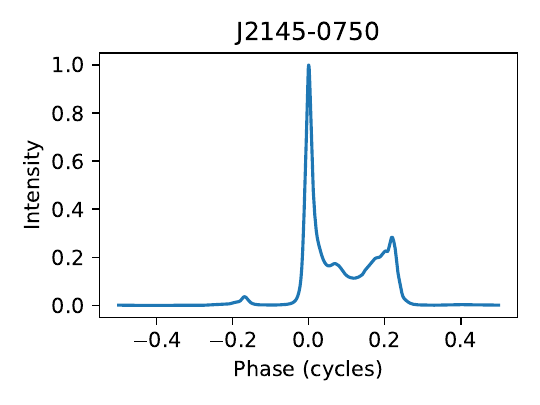}
\includegraphics[width=0.33\textwidth]{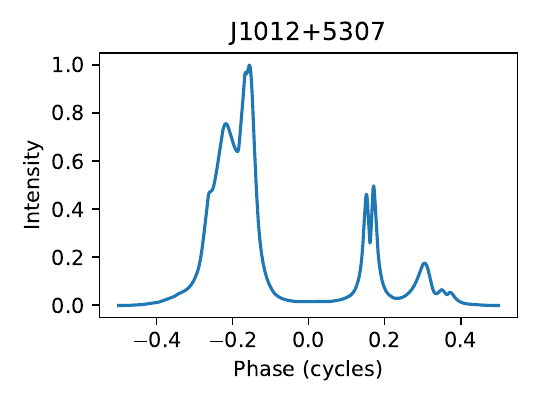}
\caption{
Average profiles for three millisecond pulsars (PSRs J1909$-$3744, J2145$-$0750, and J1012$+$5307), demonstrating the variety of pulse shapes exhibited by MSPs. All three profiles were taken from the NANOGrav 12.5-year data release \citep{12yr-nb-timing}, and are derived from observations made using the 1--2 GHz receiver and GUPPI backend on the Green Bank Telescope between 2010 and 2017.}
\label{fig:msp-profiles}
\end{figure}

All applications of pulsar timing, including those given above, rely on the ability to construct precise and accurate TOA estimates. Current state-of-the-art methods for estimating TOAs from radio observations of pulsars use template-matching algorithms, most commonly that of \citet{taylor92}, which uses Fourier-domain cross-correlation (see section~\ref{sec:toa-estimation}) to fit a model of the form
\begin{equation}\label{eqn:basic-model}
p(\phi) = a\mkern2mu u(\phi-\tau) + b + n(\phi)
\end{equation}
to the observed pulse profile, $p(\phi)$, that is, the phase-resolved total intensity of radio emission from the pulsar, accumulated and folded into a finite number, $N_\phi$, of phase bins, using a pre-computed ephemeris. In equation~(\ref{eqn:basic-model}), $\phi$ is pulse phase, measured in turns; $u(\phi)$ is a template; $a$ is a scale parameter or amplitude, $b$ is a DC offset or bias; $\tau$ is the phase shift to be measured; and $n(\phi)$ represents white, Gaussian noise. The template, $u(\phi)$, is constructed as a representation of the long-term average profile shape. This average shape is not nessarily simple. Especially in the case of MSPs, it can often be often complex and multi-peaked, as seen in Figure~\ref{fig:msp-profiles}. To produce an accurate TOA, the template must accurately reflect this sometimes complex shape.

More sophisticated template-matching algorithms expand on the basic idea by including more data in the fit: for example, matrix template matching \citep{van-straten06} models the Stokes parameters $Q$, $U$, and $V$, along with the total intensity, $I$; and wideband timing \citep{pdr14} models the profile across a range of frequency channels simultaneously. But the basic principle remains the same: one or more observed profiles are fit using a template which is shifted by an unknown phase offset, $\tau$, and the recovered phase offset, $\hat\tau$, is used to produce a TOA estimate.\footnote{This is done by scaling $\hat\tau$ by the pulse period to produce a time offset, and adding the result to the time closest to the center of the observation which corresponds to a whole number of turns according to the folding ephemeris.}

Template-matching algorithms work well as long as the profile matches the template; in other words, as long as the pulse profile has a known, fixed shape, modified only by an overall amplitude, bias, and phase shift, and by the addition of white noise, as described in equation~(\ref{eqn:basic-model}). This is usually a good approximation \citep[cf.][]{hmt75}, but it has important limitations. In particular, single pulses can have very different shapes, and can occur at different pulse phases \citep{craft70,jak+98,es03}. These short-term variations in pulse shape mean that observed profiles do not, in general, exactly match the template. This results in errors in the resulting TOA estimates, a phenomenon known as jitter noise \citep{cd85,sod+14}.

The fact that jitter noise arises from a mismatch between the observed profile and the template, i.e., a failure of the simple fixed-template model~(\ref{eqn:basic-model}) to correctly describe the data, suggests that improved models may be able to mitigate its effects. This idea has motivated several attempts to improve pulsar timing by incorporating additional information about the profile shape. The earliest attempt of this type appears to have been that of \citet{dkp85}, who applied a correction method based on a shape parameter, along the lines we describe in Section~\ref{sec:correction-general}, with limited success, achieving a $\sim$ 30\% correction for only two of the 21 pulsars they analyzed. A retrospective analysis of the same data by JMC has shown that the lack of dedispersion, which produced TOA errors larger than those caused by jitter alone, was a major limiting factor in this early study.

A method based on principal component analysis (PCA), which we describe in Section~\ref{sec:pca-method}, was first introduced by \citet{demorest-thesis}, and further developed by \citet{ovh+11}, who emphasized the use of multiple regression to develop a correction incorporating information from more than one principal component at a time, and used it to achieve a $\sim$ 37\% reduction in reduced chi-squared when fitting a timing model to observations of PSR J0437$-$4715 made with Murriyang, the 64-meter telescope at Parkes Observatory. 
\citet{lah15} devised a related scheme, using a basis of Hermite functions referred to as ``shapelets'', as part of a broader program in which profile shapes were inferred simultaneously with the timing model. This ``generative pulsar timing analysis'' has the potential to be very computationally expensive for large data sets, and as a result has not, to date, been widely adopted. \citet{kerr15}, working with data from the Vela pulsar, PSR B0833$-$45, used a clustering technique to produce a a representative set of single pulses, and modeled profiles as linear combinations of these. More recently, \citet{nrm+23} and \citet{slm25} have explored fitting models with multiple Gaussian components as an approach to correcting for shape variation.

In this paper, we introduce a new technique for mitigating errors caused by pulse shape variation, which extends the traditional template matching algorithm, and compare it to two other techniques: the principal component dot product method used by \citet{demorest-thesis} and \citet{ovh+11}, and an approach based on the profile skewness function, a shift-invariant measure of pulse asymmetry.

\section{TOA Estimation and Errors from White Noise}
\label{sec:toa-estimation}

\noindent The conventional template-matching algorithm used for TOA estimation is laid out in detail in \citet[Appendix A]{taylor92}. It can be characterized as the maximum likelihood estimator for the model in which the profile, $p(\phi)$, is described as a shifted, scaled copy of a fixed template, $u(\phi)$, with additive noise:
\begin{equation}\label{eqn:profile-model-cp}
    p(\phi)=a\mkern2mu u(\phi-\tau) + b + n(\phi).
\end{equation}
In the equation above, $u(\phi)$ is the template (normalized to unit maximum), $n(\phi)$ is white, Gaussian noise with variance $\sigma_n^2$, and the amplitude $a$ and phase offset $\tau$ should be interpreted as model parameters. Both $p(\phi)$ and $u(\phi)$ are assumed to be periodic in $\phi$, with unit period (i.e., we adopt the convention that the phase, $\phi$, is measured in cycles, rather than angular units). We are using a continuous notation for clarity here. In reality, values of $p(\phi)$ and $u(\phi)$ are known only at a finite number, $N_\phi$, of phase bins, corresponding to evenly sampled values of the phase, $\phi_i$, and phase shifts can be accomplished by transforming to the frequency domain using the discrete Fourier transform (DFT), multiplying by the appropriate phase factor, and transforming back using the inverse DFT. This is equivalent to constructing a continuous interpolating function (see Appendix~\ref{app:ml-ab} for additional detail).

The log-likelihood for the model given by equation~(\ref{eqn:profile-model-cp}) takes the form
\begin{equation}\begin{split}\label{eqn:loglike-cp}
    \log\mc{L}_{\mr{TM}}\of{a,b,\tau} &= -\frac{N_\phi}{2\sigma_n^2}\int_0^1 \sbrack{p(\phi) - a\mkern2mu u(\phi-\tau) - b}^2 \dd\phi - \frac{N_\phi}2\log\of{2\pi\sigma_n^2}.
\end{split}\end{equation}
For fixed $\tau$, this is a standard linear regression problem for the parameters $a$ and $b$, so the maximum likelihood values $\hat{a}(\tau)$ and $\hat{b}(\tau)$ can be determined analytically. In particular, we have
\begin{equation}\label{eqn:ab-hat}
\begin{split}
    \hat{a}(\tau) = \frac{C(\tau) - \bar{u}\mkern2mu\bar{p}}{\overline{u^2}-\bar{u}^2},\quad
    \hat{b}(\tau) = \frac{\overline{u^2}\mkern1mu\bar{p} - \bar{u}\mkern2mu C(\tau)}{\overline{u^2}-\bar{u}^2},
\end{split}
\end{equation}
where overlined quantities represent averages in pulse phase, and
\begin{equation}\label{eqn:corr-fn}
C(\tau) = \int_0^1 p(\phi)\mkern2mu u(\phi-\tau)\dd\phi
\end{equation}
is the cross-correlation between the profile and the template.
A derivation of these results can be found in  Appendix~\ref{app:ml-ab}.

Setting $a$ and $b$ to their respective maximum likelihood values, we arrive at an expression for the log-likelihood:
\begin{equation}
    \log\mc{L}_{\mr{TM}}^{(\mr{max})}\of[big]{\tau}
    = -\frac{N_\phi}{2\sigma_n^2}\curly{\overline{p^2} - \bar{p}^2 - \frac{\sbrack{C(\tau)-\bar{u}\mkern2mu\bar{p}}^2}{\overline{u^2}-\bar{u}^2}} - \frac{N_\phi}2\log\of{2\pi\sigma_n^2},
\end{equation}
It follows that the maximum likelihood value of $\tau$ is
\begin{equation}\label{eqn:max-problem}
    \hat{\tau}=\operatorname*{argmax}_\tau\,\sbrack{C(\tau)-\bar{u}\mkern2mu\bar{p}}^2 = \operatorname*{argmax}_\tau C(\tau).
\end{equation}
The last equality holds as long as $C(\hat\tau) > \bar{u}\mkern2mu\bar{p}$, which is ordinarily the case.\footnote{In principle, we could have $C(\hat\tau) < \bar{u}\mkern2mu\bar{p}$, but this would mean that $\hat{a}(\hat\tau)<0$; in other words, that the best-fit model describes the profile as containing an inverted copy of the template. In the pulsar timing context, this does not make sense on physical grounds (it would correspond to a negative value of the flux or antenna gain). However, in some very low signal-to-noise cases, such an inverted model may nevertheless be the best fit. In such cases, maximizing the likelihood for the model~(\ref{eqn:profile-model-cp}) instead corresponds to \emph{minimizing} the cross-correlation, $C(\tau)$.} The time-of-arrival estimate, $\hat\tau$, that maximizes the likelihood of the model~(\ref{eqn:profile-model-cp}) can therefore be computed by maximizing $C(\tau)$ numerically, as described in \citet{taylor92}.

The uncertainties in $a$, $b$, and $\tau$ can be determined by expanding $\chi^2(a,b,\tau)$ in a power series around its minimum, as detailed in Appendix~\ref{app:ml-ab}. In particular, if the profile is well described by the model (equation~\ref{eqn:profile-model-cp}), the uncertainty in $\tau$ is given by
\begin{equation}\label{eqn:err-from-weff}
    \sigma_\tau^2=\frac{\sigma_n^2 W_{\mr{eff}}^2}{\hat{a}(\hat\tau)^2 N_\phi},
\end{equation}
where the quantity
\begin{equation}\label{eqn:weff}
    W_{\mr{eff}}=\sbrack{\int_0^1 u'(\phi_i)^2 \dd\phi}^{-1/2}
\end{equation}
is the effective pulse width.

\section{TOA errors from profile residuals}
\label{sec:jitter-noise}




\noindent Suppose that the profile, $p(\phi)$, differs from a model of the form~(\ref{eqn:profile-model-cp}) by a small amount. That is,
\begin{equation}\label{eqn:residual}
  p(\phi) = a\mkern2mu u(\phi-\tau) + b + r(\phi),
\end{equation}
where the profile residual, $r(\phi)$, is much smaller in magnitude than $a\mkern2mu u(\phi-\tau)$.  The presence of $r(\phi)$ means that the conventional template-matching estimate of the phase shift, $\hat\tau$, will differ from the true phase shift by a small amount, $\delta\tau=\hat\tau-\tau$. As long as $\delta\tau$ is small compared to the width of the template, the maximization in equation~(\ref{eqn:max-problem}) can be carried out analytically, by expanding the condition $C'\of{\tau+\delta\tau} = 0$ to first order in $\delta\tau$ and solving for $\delta\tau$. This gives
\begin{equation}\label{eqn:delta-tau-basic}
    \delta\tau = \frac{\int_0^1 p(\phi)\mkern2mu u'(\phi-\tau)\dd\phi}{\int_0^1 p(\phi)\mkern2mu u''(\phi-\tau)\dd\phi}
\end{equation}
(cf.~\citealt{demorest-thesis}, equation 2.7; \citealt{jmr+21}, equation A4). Making use of equation~(\ref{eqn:residual}) and removing terms which can be shown to vanish using integration by parts, we can rewrite this as
\begin{equation}\label{eqn:delta-tau-expanded}
    \delta\tau = \frac{\int_0^1  r(\phi)\mkern2mu u'(\phi-\tau)\dd\phi}{a\int_0^1 u\of{\phi}\mkern2mu u''\of{\phi}\dd\phi + \int_0^1  r(\phi)\mkern2mu u''(\phi-\tau)\dd\phi}.
\end{equation}
Since we are assuming that $r\of\phi\ll a\mkern2mu u\of{\phi-\tau}$, the second term in the denominator can be neglected compared to the first. After integrating by parts in the denominator, the above expression for $\delta\tau$ simplifies to
\begin{equation}\label{eqn:projection}
    \delta\tau = -\frac{\int_0^1 r(\phi) \mkern2mu u'(\phi-\tau)\dd\phi}{a\int_0^1 u'(\phi)^2\dd\phi}.
\end{equation}
In other words, the error in the conventional template-matching estimate of $\tau$ is proportional to the projection of $r(\phi)$ onto $u'(\phi-\tau)$.

Given a statistical description of the profile residual, $r(\phi)$, equation~(\ref{eqn:projection}) can be used to predict the variance of the corresponding TOA errors. In particular, we have
\begin{equation}\label{eqn:projection-variance}
    \sigma_{\hat\tau}^2 = \bracket{\delta\tau^2} = \frac{\int_0^1\int_0^1 u'(\phi-\tau) \mkern2mu u'(\phi'-\tau)\mkern2mu \bracket{r(\phi)\mkern2mu r(\phi')}\dd\phi\dd{\phi'}}{a^2\sbrack{\int_0^1 u'(\phi)^2\dd\phi}^2}.
\end{equation}
Notice that, when the residual is simply white noise (i.e., $\bracket{r(\phi)\mkern2mu r(\phi')} = \paren*{\sigma_n^2/N_\phi}\,\delta\of{\phi-\phi'}$), equation~(\ref{eqn:projection-variance}) reduces to
\begin{equation}\label{eqn:predictive-err-from-weff}
\sigma_{\hat\tau}^2 = \frac{\sigma_n^2}{a^2\sbrack{\int_0^1 u'(\phi)^2\dd\phi}} = \frac{\sigma_n^2 W_{\mr{eff}}^2}{a^2 N_\phi},
\end{equation}
where $W_{\mr{eff}}$ is given by equation~(\ref{eqn:weff}). Equation~(\ref{eqn:predictive-err-from-weff}) is very similar to equation~(\ref{eqn:err-from-weff}), but it has a slightly different interpretation: equation~(\ref{eqn:predictive-err-from-weff}) gives the mean squared error in the TOA estimate, $\hat\tau$, for a specified underlying model, while equation~(\ref{eqn:err-from-weff}) gives the variance of the posterior distribution for the underlying TOA, $\tau$, considered as a model parameter, for specified observed data (assuming a prior distribution that is uniform in $\tau$).

If the profile is considered to be made up of several Gaussian components with stochastically varying amplitude and phase, equation~(\ref{eqn:predictive-err-from-weff}) can be used to make detailed predictions about the variance of the resulting TOA errors. This is explored in detail in Appendix~\ref{sec:multi-component-characterization}.

It may be tempting to try to correct TOA estimates using equation~(\ref{eqn:projection}) directly. For example, one might consider fitting a model of the form~(\ref{eqn:profile-model-cp}) to the profile to obtain an initial TOA estimate, $\hat\tau$, and estimates of the amplitude parameters, $\hat{a}(\hat\tau)$ and $\hat{b}(\hat\tau)$, then estimating $r(\phi)$ by subtracting this best-fit model from the profile, and finally estimating $\delta\tau$ using equation~(\ref{eqn:projection}) (replacing $a$ with $\hat{a}(\hat\tau)$) and subtracting it from $\hat\tau$. However, this is not a useful approach. The reason for this was understood already by~\citet[][Section 2.3]{demorest-thesis}. In particular, note that by following such an approach, one would arrive at the following estimate of $r(\phi)$:
\begin{equation}\label{eqn:estimated-residual}
\begin{split}
    \hat{r}(\phi) &= p(\phi) - \hat{a}(\hat\tau)\mkern2mu u(\phi-\hat\tau) - \hat{b}(\hat\tau).\\
    &= r(\phi) + a\mkern2mu u(\phi-\tau) - \hat{a}(\hat\tau)\mkern2mu u(\phi-\hat\tau) + b - \hat{b}(\hat\tau).
\end{split}
\end{equation}
After expanding to first order in $\delta\tau = \hat\tau - \tau$, $\delta{a}=\hat{a}(\hat\tau)-a$, and $\delta{b}=\hat{b}(\hat\tau)-b$, equation~(\ref{eqn:estimated-residual}) becomes
\begin{equation}\label{eqn:hidden-deltatau}
    \hat{r}(\phi) = r(\phi) + a\,\delta\tau\,u'(\phi-\tau) - \delta{a}\,u(\phi-\tau)-\delta{b}.
\end{equation}
Substituting equation~(\ref{eqn:hidden-deltatau}) into equation~(\ref{eqn:projection}), we find
\begin{equation}
	\delta{\hat\tau} = \frac{\int_0^1 r(\phi) \mkern2mu u'(\phi-\tau)\dd\phi - a\,\delta\tau\int_0^1 u'(\phi)^2\dd\phi}{a\int_0^1 u'(\phi)^2\dd\phi} = \delta\tau - \delta\tau = 0.
\end{equation}
(The terms involving $\delta a$ and $\delta b$ can be shown to vanish using integration by parts.) This is only a first-order calculation, so in practice, $\delta\hat\tau$ will not be exactly zero, but because of the cancellation, $\delta\hat\tau$ cannot be taken as a reasonable estimate of $\delta\tau$. In other words, because of the way $\hat\tau$ is used in producing it, the estimated residual, $\hat{r}(\phi)$, is naturally biased in a way that makes it useless for predicting $\delta\tau$ by means of equation~(\ref{eqn:projection}). This means that equation~(\ref{eqn:projection}) cannot be used to correct TOA estimates directly. Instead, other approaches are needed.

\section{Jitter correction methods}
\label{sec:correction}

In this section, we will describe methods which, unlike those described in the previous section, can be used to produce corrected TOA estimates for data affected by shape variations. We will assume that these shape variations arise from jitter, i.e., that the profile can be described as
\begin{equation}\label{eqn:profile-model-jitter}
p(\phi) = a\sbrack{u(\phi-\tau) + \delta u(\phi-\tau)} + b + n(\phi),
\end{equation}
where $\delta u(\phi)$ is a shape perturbation satisfying $\bracket{\delta u(\phi)} = 0$ and 
\begin{equation}
\bracket{\delta u(\phi)\mkern2mu \delta u(\phi')} = D(\phi, \phi').
\end{equation}
As in equation~(\ref{eqn:profile-model-cp}), $u(\phi)$ is the template (average profile shape), $a$ and $b$ are amplitude parameters, and $n(\phi)$ is noise, with variance $\sigma_n^2$ in each phase bin. Comparing equations~(\ref{eqn:residual}) and~(\ref{eqn:profile-model-jitter}) shows that the profile residual is given by
\begin{equation}
r(\phi) = a\mkern2mu\delta u(\phi-\tau) + n(\phi).
\end{equation}
The covariance matrix of $r(\phi)$ is therefore
\begin{equation}\label{eqn:resid-covmat}
R(\phi,\phi') = \bracket{r(\phi)\mkern2mu r(\phi')} = a^2 D(\phi,\phi') + \frac{\sigma_n^2}{N_\phi}\mkern2mu\delta(\phi-\phi').
\end{equation}
Importantly, we also assume that the shape perturbations $\delta u(\phi)$ are identically distributed for successive pulses, i.e., that they are stationary in time. If this assumption holds, it follows that the covariance matrix $D(\phi,\phi')$ is time-independent. The assumption of stationarity appears to be realistic for pulsar data, although it has not been extensively tested. On the other hand, it cannot, in general, be assumed that successive pulses are \emph{independent}: phenomena such as subpulse drifting~\citep{dc68,rankin86,sws+23} and deviations from $\sqrt{N}$ averaging behavior~\citep{hmt75,lkl+12} demonstrate that, at least in some cases, correlations exist between successive pulses. It is possible that the methods described below can be improved by accounting for these pulse-to-pulse correlations, but here, we note only that stationarity still seems to hold, even in cases where the shapes of successive pulses are clearly not independent.

In this setting, there are several possible methods that can be used to produce TOA estimates which are more accurate than those produced by the conventional template-matching algorithm. In Section~\ref{sec:correction-general}, we outline some general considerations that apply to any such method. In section~\ref{sec:skewness-method}, we describe a simple correction method based on the profile skewenss function. In section~\ref{sec:pca-method}, we introduce a correction method based on principal component analysis, used previously by \citet{demorest-thesis} and \citet{ovh+11}. Finally, in section~\ref{sec:gen-template-matching}, we introduce our new generalized template-matching algorithm, and compare it to the previously described methods.

\subsection{General considerations}
\label{sec:correction-general}

It is possible to derive a correction method from any measure of pulse shape which is correlated with the TOA error. Suppose $x\sbrack{p(\phi)}$ is a function of the profile shape (a ``shape parameter'') satisfying $x\sbrack{a\mkern2mu u(\phi)}=0$ for any $a$, and that the profile residual, $r(\phi)$, is drawn from a consistent distribution with mean zero. We can then calculate the correlation coefficient, $\rho$, between $x$ and $\delta\tau=\tau-\hat\tau_0$, where $\hat\tau_0$ is the uncorrected, fixed-template TOA estimate:
\begin{equation}
    \rho = \frac{\bracket{x\mkern2mu\delta\tau}}{\sqrt{\bracket{x^2}\bracket{\delta\tau^2}}},
\end{equation}
where angle brackets denote ensemble averages. If $\rho$ is nonzero, we can exploit the correlation to construct a new, improved estimate of $\tau$. Among estimators of the form $\hat\tau = \hat\tau_0 + c x$, the one minimizing the mean squared error is
\begin{equation}\label{eqn:shape-parameter-correction}
    \hat\tau = \hat\tau_0 - \rho\sqrt{\frac{\bracket{\delta\tau^2}}{\bracket{x^2}}}x.
\end{equation}
The remaining mean squared error is given by
\begin{equation}
    \bracket{(\hat\tau-\tau)^2} = (1-\rho^2)\bracket{\delta\tau^2}.
\end{equation}
This is always less than the mean squared error of the uncorrected estimate, $\bracket{(\hat\tau_0-\tau)^2}=\bracket{\delta\tau^2}$. The correction is most effective when $\rho^2\approx 1$, i.e., when $x$ and $\delta\tau$ are highly correlated.

An important consideration with any TOA estimation method based on profile shape is that it is difficult to completely separate the notion of the shape of a profile from its location in pulse phase. If this is not handled carefully, the result can be a correction scheme that removes any deviation from prior expectations, including the delays which were intended to be measured. To avoid this problem, it is desireable for the TOA estimate, $\hat\tau$, to have the property
\begin{equation}\label{eqn:shift-covariance}
    \hat\tau\sbrack{p(\phi-\delta\phi)} = \hat\tau\sbrack{p(\phi)} + \delta\phi.
\end{equation}
Equation~(\ref{eqn:shift-covariance}) encodes the requirement that a shift in the profile should produce a corresponding shift in the TOA estimate. This guarantees that phase shifts that affect the entire profile, such as those caused by gravitational waves or other timing signals of interest, are not removed or reduced in amplitude by the TOA estimation method. If equation~(\ref{eqn:shift-covariance}) holds for all possible values of the shift, $\delta\phi$, for a specific TOA estimation method, we say that the method is shift-covariant. In particular, the fixed-template TOA estimate,
\begin{equation}\label{eqn:tau0}
    \hat\tau_0 = \operatorname*{argmax}_\tau \int_0^1 p(\phi)\mkern2mu u(\phi-\tau)\dd\phi,
\end{equation}
described in Section~\ref{sec:toa-estimation} is shift-covariant: replacing $p(\phi)$ with $p(\phi-\delta\phi)$ in equation~(\ref{eqn:tau0}) makes the function to be maximized into
\begin{equation}
\int_0^1 p(\phi-\delta\phi)\mkern2mu u(\phi-\tau)\dd\phi = \int_0^1 p(\phi)\mkern2mu u(\phi-\tau+\delta\phi)\dd\phi,
\end{equation}
where the two sides are equivalent by a change of variables. It follows that, if the original function was maximized for $\tau=\hat\tau_0$, the new function will be maximized for $\tau-\delta\phi=\hat\tau_0$, i.e., for $\tau=\hat\tau_0+\delta\phi$, so equation~(\ref{eqn:shift-covariance}) holds.

Unlike the fixed-template TOA estimate, TOA estimates including a shape parameter-based correction are not necessarily shift-covariant.
From equation~(\ref{eqn:shape-parameter-correction}) and the fact that the fixed-template method is shift-covariant, it follows that a TOA estimation method using shape parameter-based correction satisfies
\begin{equation}
\hat\tau\sbrack{p(\phi-\delta\phi)} = \hat\tau\sbrack{p(\phi)} + \delta\phi - \rho\sqrt{\frac{\bracket{\delta\tau^2}}{\bracket{x^2}}}\curly{x\sbrack{p(\phi-\delta\phi)} - x\sbrack{p(\phi)}}.
\end{equation}
In other words, such a method will be shift-covariant only if (and only if) the underlying shape parameter is shift-\emph{invariant}, satisfying
\begin{equation}\label{eqn:shift-invariance}
    x\sbrack{p(\phi)} = x\sbrack{p(\phi-\delta\phi)}
\end{equation}
for any shift, $\delta\phi$. Below, we will develop a TOA correction method based on the skewness function, which is an example of just such a shift-invariant shape parameter.

\subsection{The skewness method}
\label{sec:skewness-method}

One relatively easy way to produce shift-invariant shape parameters is via the autocorrelation function (ACF):
\begin{equation}
    R(\tau) = \frac1{\overline{p^2}}\int_0^1 p(\phi)\mkern2mu p(\phi+\tau) \dd{\phi}.
\end{equation}
A simple change of variables, using the assumption that $p(\phi)$ is periodic with unit period, shows that $R(\tau)$ is shift-invariant. Unfortunately, in addition to being insensitive to shifts, $R(\phi)$ is insensitive to profile asymmetries, giving the same result for the symmetrized profile $p_s(\phi)=\frac12\sbrack{p(\phi) + p(-\phi)}$ as for $p(\phi)$. In practice, pulse asymmetries are some of the best predictors of errors in template-matching TOA estimates --- asymmetric profile residuals tend to ``pull'' the estimate to one side or another, while for symmetric profile residuals, the effects of each side tend to approximately cancel out. This can be seen quantitatively by applying equation~(\ref{eqn:projection}) to a symmetric, or nearly symmetric, profile.\footnote{Realistic profiles are not perfectly symmetric, but they typically contain at least one peak which can be roughly approximated by a symmetric model.} This means that $R(\phi)$ and statistics derived from it are not particularly useful as shape parameters.

However, there is a related function which does capture information about pulse asymmetry, while remaining insensitive to shifts. This is the profile skewness function~\citep[cf.][]{sc81}, which is the antisymmetric part of the cross-correlation function between the profile, $p(\phi)$, and its square, $p(\phi)^2$:
\begin{equation}\label{eqn:skewness-fn}
    K(\tau) = \frac1{\overline{p^3}}\int_0^1 \sbrack{p(\phi)^2\mkern2mu p(\phi+\tau) - p(\phi)\mkern2mu p(\phi+\tau)^2}\dd\phi.
\end{equation}
As with $R(\tau)$ above, a change of variables, making use of the assumption that $p(\phi)$ is periodic with unit period, shows that $K(\tau)$ is shift-invariant. This means that, for any fixed value of $\tau$, if we take $x = K(\tau)$ as a shape parameter and use it to compute TOA corrections along the lines laid out in Section~\ref{sec:correction-general} above, the resulting TOA estimation method will be shift-covariant. Although the justification for them is somewhat \emph{ad hoc}, methods based on the skewness function can do surprisingly well at reducing TOA estimation errors caused by profile shape variations, as we will see in the comparisons below.


\subsection{The PCA method}
\label{sec:pca-method}

A more direct way to quantify the shape of pulses is to use principal component analysis (PCA). PCA finds the set of orthogonal modes that best capture the variance in a particular set of data. In this case, the data are the shape perturbations, $\delta u(\phi)$. PCA amounts to an eigenvalue decomposition of the covariance matrix, $D(\phi,\phi')=\bracket{\delta u(\phi)\mkern2mu \delta u(\phi')}$, of the profile residuals; i.e., we can write
\begin{equation}
D(\phi,\phi') = \sum_{k=1}^K \sigma_k^2\mkern2mu v_k(\phi)\mkern2mu v_k(\phi'),
\end{equation}
where $v_k(\phi)$ are the orthonormal principal components, $\sigma_k^2$ are the corresponding eigenvalues, and $K$ is the rank of the matrix $D$ (while $D$ is written here as a continuous function of two variables, in practice it is an $N_\phi\times N_\phi$ matrix, so $K$ is at most $N_\phi$). Since the principal components, $v_k(\phi)$, are orthonormal and span the space of shape perturbations, any particular shape perturbation can be written in the form
\begin{equation}\label{eqn:pc-decomposition}
\delta u(\phi) = \sum_{i=1}^{K} x_i\mkern2mu v_i(\phi),
\end{equation}
where the coefficients $x_i$, which we will refer to as principal component amplitudes, are given by
\begin{equation}\label{eqn:pc-shape-param}
x_i = \int_0^1 v_i(\phi)\mkern2mu \delta u(\phi)\dd\phi.
\end{equation}
By construction, the principal component amplitudes $x_i$ also satisfy $\bracket{x_i\mkern2mu x_j}=\sigma_i^2\mkern2mu\delta_{ij}$.

One can use the first $n$ principal component amplitudes jointly to construct a correction term along the lines of that described in Section~\ref{sec:correction-general}, so that the corrected TOA estimate is given by
\begin{equation}\label{eqn:pca-corrected-toa}
    \hat\tau_{\mr{PCA}} = \hat\tau_0 + \sum_{i=1}^n c_i\mkern2mu x_i,
\end{equation}
where $\hat\tau_0$ is the TOA estimate based on the conventional template-matching algorithm.
The coefficients $c_i$ which maximize the mean squared error of the corrected TOA estimate, $\bracket{(\hat\tau_{\mr{PCA}}-\tau)^2}$, are given by
\begin{equation}
    c_i = -\frac{\bracket{x_i\mkern2mu\delta\tau}}{\sigma_i^2},
\end{equation}
where, as in Section~\ref{sec:correction-general}, $\delta\tau = \hat\tau_0 - \tau$ is the error in the uncorrected estimate. This is essentially the correction method used by \citet{ovh+11} (see below for further details). A simpler version, with $n=1$, was also used earlier by \citet{demorest-thesis}.

In applying this method to real data, it is important to consider that the covariance matrix $D(\phi,\phi')$ of the shape perturbations, as well as the covariances $\bracket{x_i\mkern2mu\delta\tau}$, must be estimated using a set of ``training'' data before the method can be applied to a test data set. A simple choice is to estimate the shape perturbation for each profile, $p(\phi)$, as
\begin{equation}\label{eqn:delta-u-hat}
\delta\hat{u}(\phi) = \frac{\hat{r}(\phi+\hat{\tau})}{\hat{a}(\hat\tau)} = \frac{p(\phi+\hat\tau)-\hat{b}(\hat\tau)}{\hat{a}(\hat\tau)} - u(\phi),
\end{equation}
where $\hat\tau$ is a TOA estimate made using the conventional template-matching algorithm, with $u(\phi)$ as a template, and then to estimate $D$ as
\begin{equation}
\hat{D}(\phi,\phi') = \bracket{\delta\hat{u}(\phi)\mkern2mu\delta\hat{u}(\phi')}_t - \hat{S}_n^2\mkern2mu\delta(\phi-\phi'),
\end{equation}
where the subscript $t$ indicates that the expectation is really an average over the training data set, and $\hat{S}_n^2$ is an estimate of the contribution of noise to the diagonal.\footnote{This can be obtained in various ways, e.g., by taking an appropriately weighted mean of the sample variance of each profile within an off-pulse window, or of the high-frequency Fourier modes of each profile. Its precise value does not affect the conclusions that follow.} As recognized by \citet{ovh+11}, this estimation procedure has a surprising benefit --- it makes the method as a whole at least approximately shift-covariant.

This happens because $\hat\tau$ maximizes $C(\tau)$, the cross-correlation between the template and the profile (equation~\ref{eqn:corr-fn}), so
\begin{equation}
C'(\hat\tau) = -\int_0^1 p(\phi+\hat\tau)\mkern2mu u'(\phi)\dd\tau = 0.
\end{equation}
Using equation~(\ref{eqn:delta-u-hat}), we can therefore establish that the estimated shape perturbations, $\delta\hat{u}(\phi)$, are orthogonal to the derivative of the template, $u'(\phi)$. This orthogonality carries over to $\hat{D}(\phi,\phi')$, and therefore to the eigenvectors, $\hat{v}_k(\phi)$, derived from it; i.e., for all $k$, we have
\begin{equation}
\int_0^1 v_k(\phi)\mkern2mu u'(\phi)\dd\tau = 0.
\end{equation}
This makes the principal component amplitudes, $x_i$, approximately shift-invariant. It then follows from equation~(\ref{eqn:pca-corrected-toa}) that the corrected TOA is approximately shift-covariant. However, this approximate shift-covariance only holds for small values of the shift: the method does not exhibit exact shift-covariance. This lack of exact shift-covariance motivated the authors to search for an alternative method which did have this property, which ultimately led to the discovery of the method described in the next section.

\subsection{Generalized template matching}
\label{sec:gen-template-matching}

Both of the algorithms described above operate by correcting the estimate produced by the conventional template matching algorithm. Here we describe an alternative approach, in which we calculate the corrected TOA directly, using a generalization of the template-matching algorithm described in Section~\ref{sec:toa-estimation}. (Such a generalization can be thought of as a variable-template algorithm, as opposed to the fixed-template algorithm of Section~\ref{sec:toa-estimation}.) In particular, as we will see below, the maximum-likelihood estimate of $\tau$ for a model of the form~(\ref{eqn:profile-model-jitter}) is shift-covariant for the same reasons as the standard template-matching algorithm, and takes into account shape variations in much the same manner as the PCA-based method described in Section \ref{sec:pca-method} above. This gives it two important advantages: it is more flexible than the skewness-based method, and, unlike the PCA-based method of Section~\ref{sec:pca-method}, it is exactly shift-covariant.

The log-likelihood for the model given by equation~(\ref{eqn:profile-model-jitter}) takes the form
\begin{equation}\label{eqn:loglike-jitter}\begin{split}
\log\mc{L}(a,b,\delta u,\tau) &= -\frac{N_\phi}{2\sigma_n^2}\int_0^1 \sbrack{p(\phi) - a\mkern2mu u(\phi-\tau) - a\mkern2mu \delta u(\phi-\tau) - b}^2\dd\phi -\frac{N_\phi}2\log\of{2\pi\sigma_n^2}.
\end{split}\end{equation}
If $\delta u(\phi)$ is written in terms of principal components, $v_i(\phi)$, and the corresponding amplitudes, $x_i$, as in equation~(\ref{eqn:pc-decomposition}), this becomes
\begin{equation}\label{eqn:loglike-jitter-xi}
\log\mc{L}(a,b,x_i,\tau) = -\frac{N_\phi}{2\sigma_n^2}\int_0^1 \sbrack{p(\phi) - a\mkern2mu u(\phi-\tau) - a\mkern2mu \sum_{i=1}^{K} x_i v_i(\phi-\tau) - b}^2\dd\phi -\frac{N_\phi}2\log\of{2\pi\sigma_n^2}.
\end{equation}
For fixed $a$, $b$, and $\tau$, the above expression is quadratic in the principal component amplitudes, $x_i$, so the maximum-likelihood $x_i$ can be determined analytically. Specifically, the likelihood is maximized when $x_i=\hat{x}_i(a,\tau)$, where
\begin{equation}\label{eqn:xihat-ml}
\hat{x}_i(a,\tau) = \frac1a\int_0^1 p(\phi)\mkern2mu v_i(\phi-\tau)\dd\phi.
\end{equation}
Setting $x_i=\hat{x}_i(a,\tau)$ in equation~(\ref{eqn:loglike-jitter-xi}), we arrive at the following expression for the profile likelihood for $a$, $b$, and $\tau$:
\begin{equation}
\log\mc{L}_p\of[big]{a,b,\tau}
 = -\frac{N_\phi}{2\sigma_n^2}\int_0^1 \sbrack{p(\phi) - a\mkern2mu u(\phi-\tau) - b}^2 \dd\phi +\frac{N_\phi}{2\sigma_n^2}\sum_{i=1}^{K}\sbrack{\int_0^1 p(\phi)\mkern2mu v_i(\phi-\tau)\dd\phi}^2 - \frac{N_\phi}2\log\of{2\pi\sigma_n^2}.
\end{equation}
Except for the second term, which is indpendent of $a$ and $b$, this is identical to equation~(\ref{eqn:loglike-cp}), and so the maximum-likelihood values of $a$ and $b$, for fixed $\tau$, are the same (equation~\ref{eqn:ab-hat}). After setting $a$ and $b$ to these values, we arrive at an expression for the profile likelihood for $\tau$ alone:
\begin{equation}\label{eqn:ml-objective-fn}
\log\mc{L}_p(\tau) = -\frac{N_\phi}{2\sigma_n^2}\curly{\overline{p^2} - \bar{p}^2 - \frac{\sbrack{C(\tau)-\bar{u}\mkern2mu\bar{p}}^2}{\overline{u^2}-\bar{u}^2} - \sum_{i=1}^{K}\sbrack{\int_0^1 p(\phi)\mkern2mu v_i(\phi-\tau)\dd\phi}^2} - \frac{N_\phi}2\log\of{2\pi\sigma_n^2}.
\end{equation}
This one-dimensional function can be maximized to determine the overall maximum-likelihood value of $\tau$, $\hat\tau_{\mr{ML}}$. This estimate is shift invariant for the same reason that the fixed-template estimate, $\hat\tau_0$, is: after replacing $p(\phi)$ with $p(\phi-\delta\phi)$ in equation~(\ref{eqn:ml-objective-fn}), a change of variables in each integral (including the implicit integral in $C(\tau)$) shows that it has the same form, but with $\tau$ replaced with $\tau-\delta\phi$. Therefore, if the original function was maximized for $\tau=\hat\tau_{\mr{ML}}$, the new function will be maximized for $\tau-\delta\phi=\hat\tau_{\mr{ML}}$, i.e., for $\tau=\hat\tau_{\mr{ML}}+\delta\phi$.

In the presence of shape variability, TOA estimates calculated by maximizing equation~(\ref{eqn:ml-objective-fn}) are generally more accurate than fixed-template TOA estimates, which are calculated by maximizing $C(\tau)$. However, they are significantly less precise: their nominal errors, calculated from the second derivative of the likelihood function near the peak, are larger, and they are occasionally off by a large amount. These are symptoms of the fact that $\log\mc{L}_p(\tau)$ is usually relatively flat near its peak. This happens because the model is, in a sense, too flexible: because the principal components form a complete, or nearly complete, basis for the space of ``reasonable'' profile residuals, it is usually possible to describe a profile residual as a linear combination of principal components, even if it is significantly different from anything seen in the training data.

This problem can be fixed by incorporating prior information from the training data about the amplitudes of the principal components, which has so far been neglected. In particular, we can take the prior distribution on each principal component amplitude, $x_i$, to be normal, with mean zero and variance given by the corresponding eigenvalue, $\sigma_i^2$. The log-posterior distribution is then
\begin{equation}
\log\mc{P}\of{a,b,x_i,\tau} = \log\mc{L}\of{a,b,x_i,\tau} - \sum_{i=1}^{K}\sbrack{\frac{x_i^2}{2\sigma_i^2 a^2}-\frac12\log\of{2\pi\sigma_i^2 a^2}} + c,
\end{equation}
where $\log\mc{L}\of{a,b,x_i,\tau}$ is given by equation~(\ref{eqn:loglike-jitter-xi}), and $c$ is a normalizing constant. This is still quadratic in the principal component amplitudes, $x_i$, so it is possible to find their maximum \emph{a posteriori} values analytically. The result is
\begin{equation}\label{eqn:xihat-map}
\hat{x}_i(a,\tau) = a\paren{a^2+\frac{\sigma_n^2}{N_\phi\mkern1mu\sigma_i^2}}^{-1}\int_0^1 p(\phi)\mkern2mu v_i(\phi-\tau)\dd\phi.
\end{equation}
Comparing this with equation~(\ref{eqn:xihat-ml}) shows that the prior has the effect of shrinking the $\hat{x}_i$ estimates toward zero by a factor
\begin{equation}
f_i(a) = \paren{1+\frac{\sigma_n^2}{N_\phi\mkern1mu \sigma_i^2 a^2}}^{-1}.
\end{equation}
Setting $x_i=\hat{x}_i(a,\tau)$ then gives
\begin{equation}
\log\mc{P}(a,b,\tau) = -\frac{N_\phi}{2\sigma_n^2}\int_0^1 \sbrack{p(\phi) - a\mkern2mu u(\phi-\tau) - b}^2 \dd\phi +\frac{N_\phi}{2\sigma_n^2}\sum_{i=1}^{K} f_i(a) \sbrack{\int_0^1 p(\phi)\mkern2mu v_i(\phi-\tau)\dd\phi}^2 + c',
\end{equation}
where $c'$ is again a normalizing constant. (Marginalizing over $x_i$ leads to an equivalent expression, which differs only by an additive constant.) Here, while the dependence on $b$ remains quadratic, the dependence on $a$ has become more complicated, so it is no longer possible to reduce the problem to a one-dimensional optimization over $\tau$. The maximum \emph{a posteriori} value of $\tau$, $\hat\tau_{\mr{MAP}}$, must therefore be determined using multivariate optimization techniques. As with the maximum likelihood estimate, $\hat\tau_{\mr{ML}}$, the maximum \emph{a posteriori} estimate can be shown to be shift-covariant by a change of variables. Its uncertainty, determined from the second derivative of the log-posterior or using Markov Chain Monte Carlo (MCMC) sampling techniques, is significantly less than that of the maximum likelihood estimate, due to the regularizing effect of the prior.

\section{Application to Simulated Data}

\begin{figure}
\includegraphics[width=\textwidth]{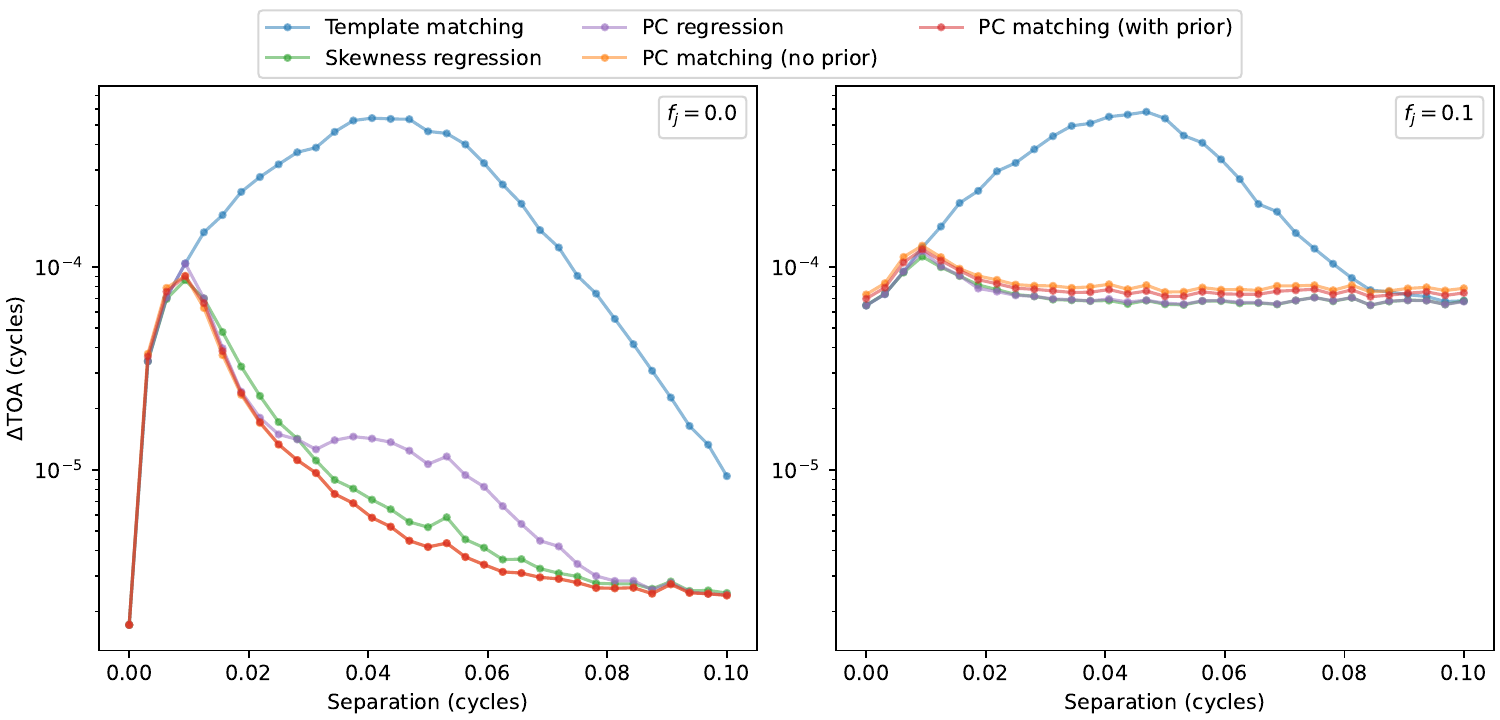}
\caption{A comparison of the results of different TOA correction methods on simulated pulse profiles. The results shown here are for simulated profiles with two identical Gaussian components, with the mean separation between the profile components shown on the horizontal axis. The standard deviation of the TOA estimates resulting from each method is shown on the vertical axis. The left panel shows the results for profiles created by averaging pulses where each component varies by 100\% in amplitude, but does not vary in its location in phase. The right panel shows the results for profiles where the center of each component varies in phase by 10\% of its width (jitter parameter $f=0.1$), in addition to exhibiting the same 100\% amplitude variations. Each point was calculated from 512 sample profiles, each of which is the average of 1000 simulated pulses. The method labeled ``PC matching'' is the new generalized template matching algorithm described in section~\ref{sec:gen-template-matching}, and that labeled ``PC regression'' is the regression-based method of \citet{ovh+11}, as described in section~\ref{sec:pca-method}. Both PCA-based methods significantly outperform the ordinary template matching method in most cases (but not all). The skewness regression method described in section~\ref{sec:skewness-method} also does remarkably well in this comparison, although it does not generalize as well.}
\label{fig:method-comparison}
\end{figure}

\begin{figure}
\includegraphics[width=\textwidth]{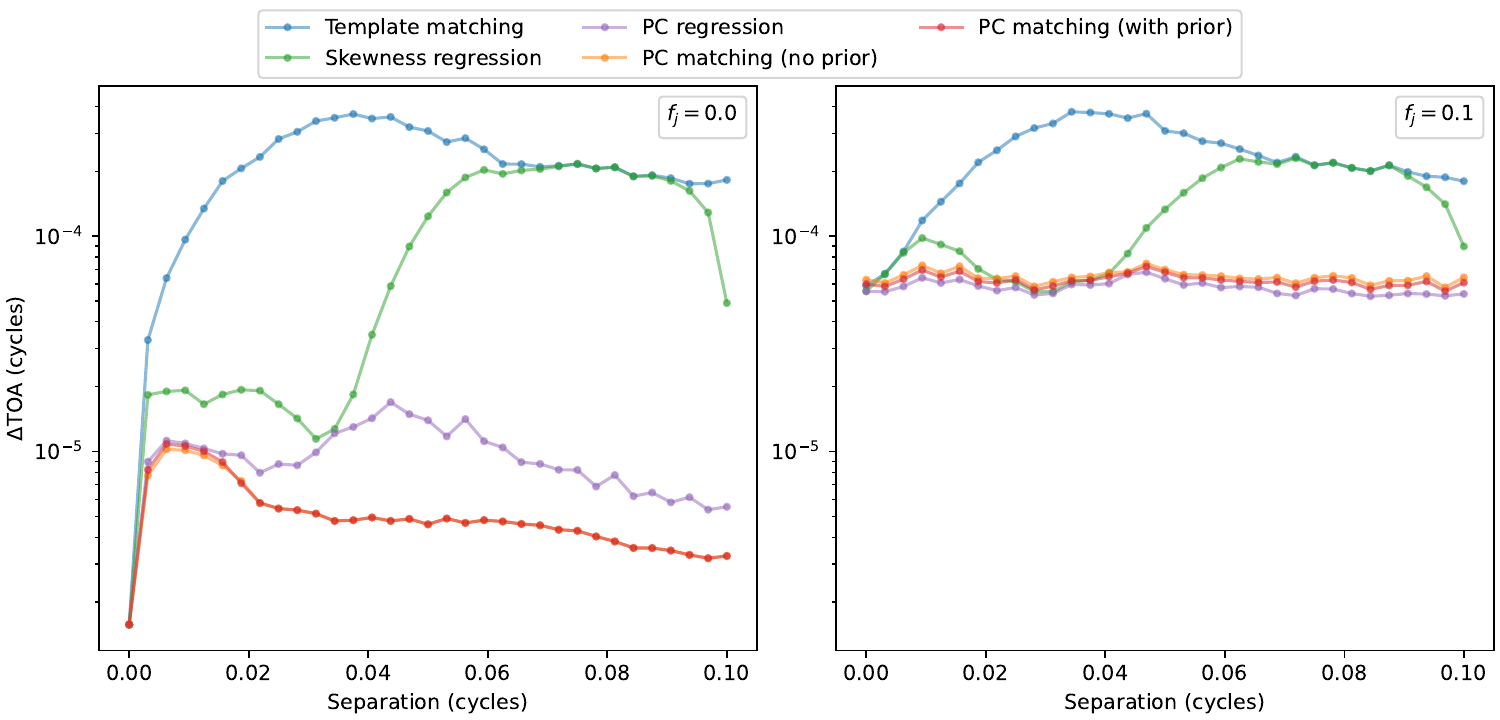}
\caption{A comparison of TOA correction methods on simulated data, as in Figure~\ref{fig:method-comparison}, but for a different test case, demonstrating that the skewness regression method, in particular, does not generalize well beyond the two-component case. In this case, the simulated profiles consist of three components, with the mean amplitude of the leading component equal to 80\% of the amplitude of the central component, and that of the trailing component is 40\% that of the central component. The separation between the leading and central components is given on the horizontal axis, while that between the central and trailing components is half of this. As in Figure 1, all three components have 100\% amplitude variations in single pulses, and the right panel differs from the left panel by the inclusion of variations in the central phase of each component equal to 10\% of the component width ($f=0.1$). Also as before, each point is derived from 512 simulated profiles, each of which is the average of 1000 simulated pulses.}
\label{fig:method-comparison-3comps}
\end{figure}

Figures~\ref{fig:method-comparison} and~\ref{fig:method-comparison-3comps} show a comparison of several correction methods, applied to simulated data. The methods used were a simple shape-parameter method based on the skewness coefficient, the principal component dot product method described in Section~\ref{sec:pca-method}, and the principal component profile reconstruction method described in Section~\ref{sec:gen-template-matching}.

The simulations are based on a model in which individual pulses are composed of several Gaussian components, each of which is subject to amplitude and phase variations. In brief, each simulated profile is given by a sum of the form
\begin{equation}
p(\phi) = \frac1N\sum_{i=1}^N\sum_{j=1}^{N_c}a_j(1+b_{ij})\exp\sbrack{-\frac{\paren*{\phi-\mu_j-\psi_{ij}}^2\ln{2}}{w_j^2}} + n(\phi),
\end{equation}
where $N$ is the number of pulses; $N_c$ is the number of components; $a_j$, $\mu_j$, and $w_j$ are the average amplitude, central phase, and full width at half maximum (FWHM) of each component, respectively; and $n(\phi)$ represents white noise. The parameters $b_{ij}$ and $\psi_{ij}$ are responsible for the amplitude and phase variations, with $\psi_{ij}$ being drawn from a normal distribution with variance $\bracket[big]{\psi_{ij}^2}=f_j^2w_j^2$, and $1 + b_{ij}$ drawn from a log-normal distribution with unit mean and variance $\bracket[big]{b_{ij}^2}=m_j^2$. The parameters $f_j$ and $m_j$ are known as the jitter parameter and modulation index, respectively.

The results shown in Figure~\ref{fig:method-comparison} are for simulations in which the pulses are composed of identical Gaussian components, each with amplitude $a=1$, FWHM $w=0.05$, and modulation index $m=1$. The separation between the centers of the components varies between simulations, and is shown on the horizontal axis. The left panel shows results for simulations with two comopnents, while the right panel shows results for simulations with three components, with an asymmetric pattern of separations. For the principal component-based methods, the number of principal components is set at $k=1$ for the two-component results, and $k=2$ for the three-component results. This is the minimum number of principal components necessary to reproduce the amplitude variations.

In many of the cases shown, the correction methods perform well. They perform particularly well in cases where the majority of the TOA estimation is the result of component amplitude variations, rather than phase variations. In the absence of phase variations, the correction methods reduce the TOA estimation error by well over 90\% in many cases. When they are present, the reduction in error using all three methods is much less, indicating that phase variations introduces a component of error that is uncorrectable, at least with these methods. It is also notable that, in the presence of phase variations, the skewness-based method outperforms the two PCA-based methods for some parameter values, when the reverse is usually the case.

\section{Summary and conclusion}
The conventional template matching algorithm for TOA estimation described by \citet{taylor92} can be described as the maximum-likelihood estimator for a model in which the pulse profile is a scaled, shifted copy of a template profile, with the addition of an unknown intensity offset, $b$, and Gaussian noise uncorrelated in pulse phase. This model can be improved to account for pulse jitter either by introducing a correction related to the TOA error by a regression model, as in the skewness-based method or the PCA-based method of \citet{ovh+11}, or by generalizing the profile model to include principal components, directly allowing for more flexible modeling of the profile shape, and using the corresponding maximum-likelihood estimator. The latter approach is our generalized template matching algorithm.

Any attempt to produce TOAs corrected for the effects of pulse jitter must take care to avoid unintentionally removing astrophysical signals of interest in the data. We argue that an estimator which is shift-covariant in the sense of equation~(\ref{eqn:shift-covariance}) will be free of such problematic effects. Maximum-likelihood methods for models with an overall shift degree of freedom, including the \citet{taylor92} algorithm and our own generalized template matching algorithm, are naturally shift-covariant, while regression methods in general are not.

Using simulated pulse profiles, we have shown that all methods described here can succeed to some degree in reducing the effects of pulse jitter on TOA measurements, and that, in many circumstances, the generalized template matching approach outperforms the others. In cases where the pulse jitter takes the form of per-component amplitude variations, jitter noise in TOAs can be almost completely eliminated. However, if per-component phase variations are involved, corretion methods are less successful, and may face fundamental limitations. No method can correct for jitter which takes the form of an overall random offset in pulse phase applied to a pulse which is otherwise consistent in shape.

Realistic pulse shape distributions are not captured perfectly either by a model with only per-component amplitude variations or by a model with only per-component phase variations, but likely fall somewhere in between, with some correction being possible. Further research is needed to determine the extent to which correction methods like those described here can improve pulsar timing precision in real data.

\begin{acknowledgments}

The authors are members of the NANOGrav collaboration, and have received financial support through the NANOGrav Physics Frontiers Center, which has been funded by the National Science Foundation (NSF) under award numbers 1430284 and 2020265.

\end{acknowledgments}

\appendix
\section{Best-fit parameter values and uncertainties in the fixed-template model}\label{app:ml-ab}

The log-likelhood for the fixed-template model~(\ref{eqn:profile-model-cp}) is given in schematic form by equation~(\ref{eqn:loglike-cp}). However, in practice, the profile and template are not continuous functions, but are known only at the values $\phi_j=j/N_\phi$, $1\leq j\leq N_\phi$. This means that the likelihood actually involves a sum, not an integral:
\begin{equation}\begin{split}\label{eqn:loglike-re}
    \log\mc{L}\of{a,b,\tau} &= -\frac1{2\sigma_n^2}\sum_{j=1}^{N_\phi} \sbrack{p(\phi_j) - a\mkern2mu u(\phi_j-\tau) - b}^2 - \frac{N_\phi}2\log\of{2\pi\sigma_n^2}.
\end{split}\end{equation}
The expression above requires evaluating $p(\phi)$ only at the values $\phi_j$, but may require evaluating $u(\phi)$ at arbitrary phase values. We solve this problem by constructing a periodic interpolating function, $u(\phi)$, using the discrete Fourier transform (DFT). In particular, we define $u(\phi_j-\tau)$ as
\begin{equation}\label{eqn:dft-interp}
    u(\phi_j-\tau) = \frac1{N_\phi}\sum_{k=-n}^{n}\tilde{u}_k\,\ee^{2\pi\ii jk/N_\phi - 2\pi\ii k\tau},
\end{equation}
where $\tilde{u}_k = \sum_{j=1}^{N_\phi} u(\phi_j)\mkern2mu\ee^{-2\pi\ii jk/N_\phi}$ is the DFT of the sequence $u(\phi_j)$, and $n=\floor{(N_\phi-1)/2}$. That is, $u(\phi_j-\tau)$ is defined as the inverse DFT of the sequence $\tilde{u}_k' = \tilde{u}_k\mkern2mu\ee^{-2\pi\ii k\tau}$, where we have made sure to use negative $k$-values in place of those larger than the Nyquist frequency.\footnote{When $N_\phi$ is even, we also set the Nyquist frequency component, $\tilde{u}_{N_\phi/2}$, to zero. This is reasonable because the template is typically a smooth function, with most of the power concentrated at relatively low frequencies, so, at worst, setting $\tilde{u}_{N_\phi/2}=0$ should not change the shape of the template appreciably.} This makes no difference when $\tau$ is an integer, but is necessary to maintain Hermitian symmetry of $\tilde{u}_k'$ (and thus real-valuedness of the interpolating function) for fractional values of $\tau$. Equation~(\ref{eqn:dft-interp}) can be used to establish an equivalence between our time-domain expressions and the frequency-domain expressions given in \citet{taylor92}.

The expression on the right-hand side of equation~(\ref{eqn:loglike-re}) is quadratic in $a$ and $b$, so it can be maximized with respect to those variables analytically. In fact, for fixed $\tau$, this is a standard linear regression problem.
This becomes easier to see if we define the profile vector $\vec{p}$ with components $p_j = p(\phi_j)$, the design matrix $\mat{X}$ with components $X_{1j}=u(\phi_j-\tau)$, $X_{2j}=1$, and the parameter vector $\vec\theta$ with components $\theta_1 = a$, $\theta_2 = b$. Using these, we can write equation~(\ref{eqn:loglike-re}) in matrix form as
\begin{equation}\label{eqn:loglike-matrix}
    \log\mc{L}\of{\vec\theta,\tau} = -\frac1{2\sigma_n^2}\paren{\vec{p}-\mat{X}\vec\theta}^\T\paren{\vec{p}-\mat{X}\vec\theta} - \frac{N_\phi}2\log\of{2\pi\sigma_n^2}.
\end{equation}
We can complete the square in $\vec\theta$ using the identity
\begin{equation}
    \paren{\vec{p}-\mat{X}\vec\theta}^\T\paren{\vec{p}-\mat{X}\vec\theta} =  \paren*{\vec\theta-\vec{\hat\theta}}^\T\mat{X}^\T\mat{X}\paren*{\vec\theta-\vec{\hat\theta}} + \vec{p}^\T\vec{p} - \vec{p}^\T\mat{X}\paren*{\mat{X}^\T\mat{X}}^{-1}\mat{X}^\T\vec{p},
\end{equation}
where
\begin{equation}\label{eqn:theta-hat}
    \vec{\hat\theta} = (\mat{X}^\T\mat{X})^{-1}\mat{X}^\T\vec{p}.
\end{equation}
Equation~(\ref{eqn:loglike-matrix}) then becomes
\begin{equation}\label{eqn:loglike-matrix-completesquare}
    \log\mc{L}\of{\vec\theta,\tau} = -\frac1{2\sigma_n^2}\sbrack{\paren*{\vec\theta-\vec{\hat\theta}}^\T\mat{X}^\T\mat{X}\paren*{\vec\theta-\vec{\hat\theta}} + \vec{p}^\T\vec{p} - \vec{p}^\T\mat{X}\paren*{\mat{X}^\T\mat{X}}^{-1}\mat{X}^\T\vec{p}} - \frac{N_\phi}2\log\of{2\pi\sigma_n^2}.
\end{equation}
The above expression makes it clear that setting $\vec\theta=\vec{\hat\theta}$ maximizes the likelihood for any given fixed value of $\tau$. At this maximum-likelihood value of $\vec\theta$, the log-likelihood reduces to
\begin{equation}\label{eqn:loglike-matrix-profile}
    \log\mc{L}(\vec{\hat\theta}(\tau),\tau) = -\frac1{2\sigma_n^2}\sbrack{\vec{p}^\T\vec{p} - \vec{p}^\T\mat{X}\paren*{\mat{X}^\T\mat{X}}^{-1}\mat{X}^\T\vec{p}} - \frac{N_\phi}2\log\of{2\pi\sigma_n^2}.
\end{equation}
In components, we have
\begin{equation}\label{eqn:xx-components}
    \mat{X}^\T\mat{X} = \sum_{j=1}^{N_\phi}\begin{bmatrix}
    u(\phi_j-\tau)^2 & u(\phi_j-\tau)\\
    u(\phi_j-\tau) & 1
    \end{bmatrix} = N_\phi \begin{bmatrix}
    \overline{u^2} & \bar{u}\\
    \bar{u} & 1
    \end{bmatrix},
\end{equation}
where overlines represent averages over the $N_\phi$ phase bins, and the shift doesn't affect either sum because of the way $u(\phi)$ is constructed (equation~\ref{eqn:dft-interp}). Similarly, 
\begin{equation}\label{eqn:xp-components}
	\mat{X}^\T\vec{p} = \sum_{j=1}^{N_\phi}\begin{bmatrix}
	u(\phi_j - \tau)\,p(\phi_j)\\
	p(\phi_j)
	\end{bmatrix} = N_\phi\begin{bmatrix}
	C(\tau)\\
	\bar{p}
	\end{bmatrix},
\end{equation}
where $C(\tau)$ is the cross-correlation between the profile and the template, as defined in equation~(\ref{eqn:corr-fn}). Substituting equations~(\ref{eqn:xx-components}) and~(\ref{eqn:xp-components}) into equation~(\ref{eqn:theta-hat}), we see that
\begin{equation}\label{eqn:theta-hat-components}
	\vec{\hat\theta}(\tau) = \begin{bmatrix}
	\hat{a}(\tau)\\
	\hat{b}(\tau)
	\end{bmatrix} = \frac1{\overline{u^2}-\bar{u}^2}\begin{bmatrix}
	C(\tau) - \bar{u}\mkern2mu\bar{p}\\
	\overline{u^2}\bar{p} - \bar{u}\mkern2mu C(\tau)
	\end{bmatrix}.
\end{equation}
Making similar substitutions in equation~(\ref{eqn:loglike-matrix-profile}) shows that
\begin{equation}\label{eqn:loglike-profile}
    \log\mc{L}\of[big]{\hat{a}(\tau),\hat{b}(\tau),\tau}= -\frac{N_\phi}{2\sigma_n^2}\curly{\overline{p^2} - \bar{p}^2 - \frac{\sbrack{C(\tau)-\bar{u}\mkern2mu\bar{p}}^2}{\overline{u^2}-\bar{u}^2}} - \frac{N_\phi}2\log\of{2\pi\sigma_n},
\end{equation}
from which it follows that the maximum likelihood value of $\tau$ is
\begin{equation}
    \hat\tau = \operatorname*{argmax}_\tau\,\sbrack{C(\tau)-\bar{u}\mkern2mu\bar{p}}^2.
\end{equation}

Expanding the log-likelihood in a power series about the maximum likelihood values, we see that
\begin{equation}
\log\mc{L}\of{\vec\theta,\tau} \approx \log\mc{L}\of*{\vec{\hat\theta}(\hat\tau),\hat\tau} - \frac1{2\sigma_n^2}\paren*{\vec\theta-\vec{\hat\theta}}^\T\mat{X}^\T\mat{X}\paren*{\vec\theta-\vec{\hat\theta}} + \frac{N_\phi}{2\sigma_n^2}\mkern2mu\hat{a}(\hat\tau)\mkern2mu C''(\hat\tau)\mkern2mu \paren*{\tau-\hat\tau}^2.
\end{equation}
It follows that, in the posterior corresponding to a flat prior in $a$, $b$, and $\tau$, the covariance matrix of $\vec\theta$ (i.e., $a$ and $b$) is given by
\begin{equation}
\bracket[big]{(\vec\theta-\vec{\hat{\theta}})(\vec\theta - \vec{\hat\theta})^\T} = \begin{bmatrix}
\sigma_a^2 & \rho_{ab}\sigma_a\sigma_b\\
\rho_{ab}\sigma_a\sigma_b & \sigma_b^2
\end{bmatrix} = \sigma_n^2\paren*{\mat{X}^\T\mat{X}}^{-1} = \frac{\sigma_n^2}{N_\phi(\overline{u^2}-\bar{u}^2)}\begin{bmatrix}
1 & -\bar{u}\\
-\bar{u} & \overline{u^2}
\end{bmatrix},
\end{equation}
and the variance of $\tau$ is approximately given by
\begin{equation}\label{eqn:sigma-tau}
\bracket{(\tau-\hat\tau)^2} = \sigma_\tau^2 = \frac{\sigma_n^2}{\hat{a}(\hat\tau)N_\phi}\mkern2mu \sbrack{-\frac1{N_\phi}\sum_{j=1}^{N_\phi} p(\phi_j)\mkern1mu u''(\phi_j - \hat\tau)}^{-1}.
\end{equation}
Using Parseval's theorem for the DFT in conjunction with equation~(\ref{eqn:dft-interp}), one can prove that
\begin{equation}
\frac1{N_\phi}\sum_{j=1}^{N_\phi} \sbrack{\hat{a}(\hat\tau)\mkern2mu u(\phi-\hat\tau) - \hat{b}(\hat\tau)}\mkern1mu u''(\phi_j - \hat\tau) = -\frac{\hat{a}(\hat\tau)}{N_\phi}\sum_{j=1}^{N_\phi} u'(\phi_j)^2,
\end{equation}
so if the profile is well-described by the model ($p(\phi_j)\approx\hat{a}(\hat\tau)\mkern2mu u(\phi_j-\tau)+\hat{b}(\hat\tau)$), the posterior variance of $\tau$ can also be written in terms of the template alone, a form which is more useful for prediction. Explicitly,
\begin{equation}
    \sigma_\tau^2 = \frac{\sigma_n^2}{\hat{a}(\hat\tau)^2 N_\phi}\mkern2mu \sbrack{\frac1{N_\phi}\sum_{j=1}^{N_\phi} u'(\phi)^2}^{-1} = \frac{\sigma_n^2 W_{\mr{eff}}^2}{\hat{a}(\hat\tau)^2 N_\phi},
\end{equation}
where
\begin{equation}\label{eqn:weff-discrete}
    W_{\mr{eff}}=\sbrack{\frac1{N_\phi}\sum_{j=1}^{N_\phi} u'(\phi)^2}^{\!-1/2}
\end{equation}
is a quantity with units of phase which can be interpreted as an effective pulse width.

\section{Predicting jitter variance in multi-component profiles}
\label{sec:multi-component-characterization}

The average profiles of pulses from pulsars, especially MSPs, are often complex and multi-peaked, as seen in Figure~\ref{fig:msp-profiles}, with components of single pulses often differing in amplitude and phase from those seen in the average profile. To model this, we can assume that the pulse is made up of several components of fixed shape, each of which varies in amplitude and phase. The profile is then given by
\begin{equation}\label{eqn:profile-components}
    p(\phi) = \frac{a}{N}\sum_{i=1}^N\sum_{j=1}^{N_c} \paren{1+b_{ij}}\mkern1mu c_{j}(\phi-\psi_{ij}),
\end{equation}
where $N$ is the number of pulses being averaged, $N_c$ is the number of components, $c_{j}(\phi)$ is the shape of component $j$, and $b_{ij}$ and $\psi_{ij}$ are the amplitude and phase offsets, respectively, of component $j$ in pulse $i$. Without loss of generality, we can take the ensemble average amplitude and phase offsets for each component to be zero. Equation~(\ref{eqn:profile-components}) can then be expanded to first order in $\psi_{ij}$, giving
\begin{equation}\label{eqn:profile-expansion}
    p(\phi) \approx \frac{a}{N}\sum_{i=1}^N\sum_{j=1}^{N_c}\sbrack{\paren{1 + b_{ij}} c_j(\phi) - \paren{\psi_{ij} + b_{ij}\psi_{ij}} c_j'(\phi)}.
\end{equation}
The template is the normalized ensemble average profile:
\begin{equation}\label{eqn:template}
    u(\phi)=\frac{\bracket{p(\phi)}}{a}\approx\sum_{j=1}^{N_c}c_j(\phi).
\end{equation}

The profile residual, $r(\phi)=p(\phi)-a\mkern2mu u(\phi)$, is therefore
\begin{equation}\label{eqn:model-residual}
    r(\phi)\approx a\sum_{j=1}^{N_c}\sbrack{\overline{b_j}\mkern2mu c_j(\phi) + \paren[big]{\overline{\psi_j} + \overline{b_j\psi_j}} c_j'(\phi)},
\end{equation}
where barred quantities represent sample averages. The first term in equation~(\ref{eqn:model-residual}) arises from the difference in amplitude between the profile and the template, and the second from the difference in phase. Substituting this into equation~(\ref{eqn:projection}) gives
\begin{equation}\label{eqn:tauhat-full}
    \delta\tau\approx\frac{\sum_{j=1}^{N_c}\sum_{k=1}^{N_c} \sbrack{\paren[big]{\overline{\psi_j} + \overline{b_j\psi_j}} \Gamma_{jk} + \overline{b_j}\mkern2mu\Delta_{jk}}} {\sum_{j=1}^{N_c}\sum_{k=1}^{N_c} \Gamma_{jk}},
\end{equation}
where we have made use of the symbols
\begin{align}
    \Gamma_{jk} &= \int_0^1  c_j'(\phi)\mkern2mu c_k'(\phi)\dd\phi\quad\text{and}\label{eqn:Gamma}\\
    \Delta_{jk} &= \int_0^1  c_j(\phi)\mkern2mu c_k'(\phi)\dd\phi\label{eqn:Delta}
\end{align}
for cross-correlations between the component shapes, and ignored all but the leading term in the denominator.
When there is only one component, equation~(\ref{eqn:tauhat-full}) reduces to
\begin{equation}
    \hat\tau\approx\overline{\psi}+\overline{b\psi}.
\end{equation}
The variance of the TOA estimate is therefore
\begin{equation}\label{eqn:toa-variance-1comp}
    \sigma_\tau^2=\bracket[big]{\hat\tau^2}=\frac{1+m^2}{N}\bracket[big]{\psi^2},
\end{equation}
where $m^2=\bracket{b^2}$ is the square of the single-pulse modulation index. This shows that, for a single-component pulse, phase variation alone can produce TOA errors. Amplitude variation may enhance TOA errors created by phase variation, but cannot produce them on its own.

For an arbitrary number of components,  the variance of the TOA estimate from equation~(\ref{eqn:tauhat-full}) is
\begin{equation}\label{eqn:tauhat-variance}
    \sigma_\tau^2=\frac{\sum_{j=1}^{N_c}\sum_{k=1}^{N_c} \sbrack{\paren{1+m_j^2}\bracket{\psi_j^2} \Gamma_{jk}^2 + m_j^2 \Delta_{jk}^2}}{N\paren{\sum_{j=1}^{N_c}\sum_{k=1}^{N_c} \Gamma_{jk}}^2},
\end{equation}
where $m_j^2=\bracket{b_j^2}$ is the square of the single-pulse modulation index for component $j$. Unlike in the single component case, here amplitude variation can create TOA errors on its own, even in the absence of phase variation: setting $\bracket{\psi_j}^2=0$ in equation~(\ref{eqn:tauhat-variance}) leaves the term involving $\Delta_j^2$. Conceptually, this is because amplitude variations represent true shape changes only when there are multiple components with separately varying amplitudes. For single component pulses, amplitude variations simply scale the pulse, and can be absorbed into the phase factor, $a$, but for pulses with multiple components that vary separately, this is no longer the case.

\subsection{Simulations with Gaussian Components}

\begin{figure}
\centering
\includegraphics[width=0.7\textwidth]{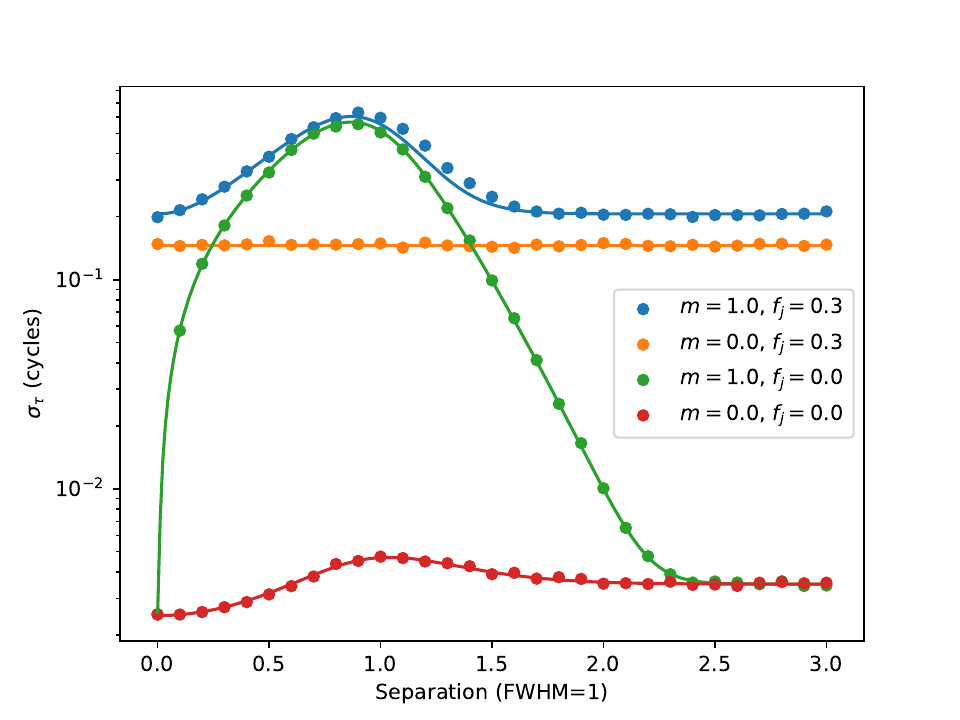}
\caption{
The standard deviation of TOA estimates in simulated profiles with two identical components of width $w=0.0425$ and varying spacing. Crosses indicate results of simulated cases, and solid lines indicate predictions based on equations~(\ref{eqn:err-from-weff}) and~(\ref{eqn:twocomp-prediction}). Four cases are shown: one in which the pulses have amplitude and phase variations ($m=1.0$, $f=0.3$ for each component), one in which the pulses have only phase variations ($m=0.0$, $f=0.3$), one in which they have only amplitude variations ($m=1.0$, $f=0.0$), and a reference case where the pulses are copies of the template with additive white noise ($m=0.0$, $f=0.0$). Each point was calculated based on $N_p=2048$ profiles, each the average of $N=1000$ pulses. The average profiles had a signal-to-noise ratio $S=1000$.}
\label{fig:2comps-sep}
\end{figure}

\begin{figure}
\centering
\includegraphics[width=\textwidth]{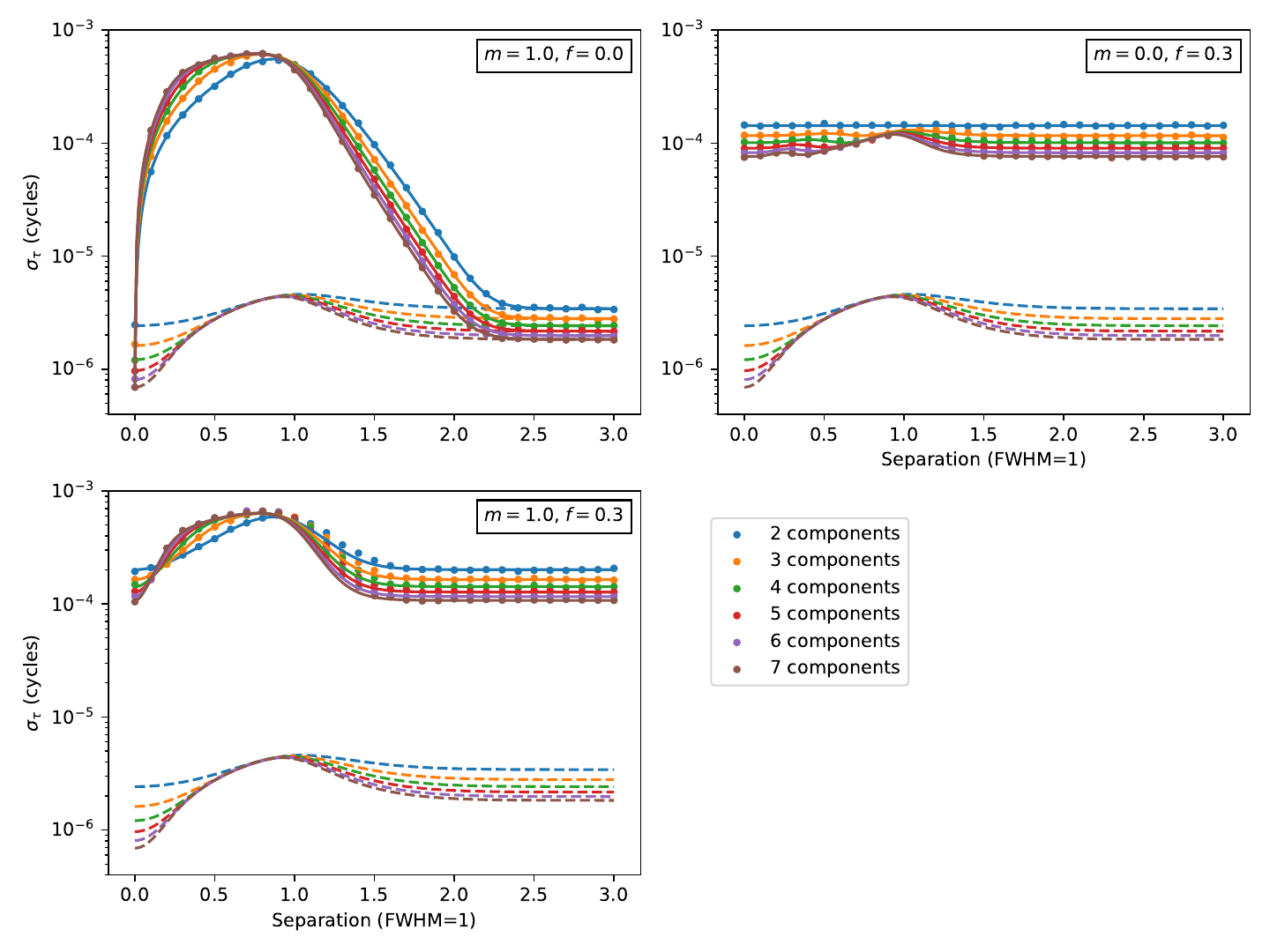}
\caption{The standard deviation of TOA estimates in simulated profiles with several identical components of width $w=0.0425$ and varying spacing. Results with amplitude variations only ($m=1.0$) is shown in the upper left, results with phase variations only ($f_j=0.3$) in the upper right, and results with both amplitude and phase variations ($m=1.0$, $f_j=0.3$) in the lower left. As in Figure~\ref{fig:2comps-sep}, filled circles indicate simulation results, and solid lines indicate predictions based on equation~(\ref{eqn:tauhat-variance}). For comparison, the predictions for the reference case with no amplitude or phase variations (based on equation~\ref{eqn:err-from-weff}) are shown as dashed lines.}
\label{fig:ncomps-sep}
\end{figure}

We performed a number of simulations in which shape variations were introduced into a simple pulse shape model consisting of Gaussian components. In each simulation, $N_p$ profiles were generated by averaging $N$ simulated pulses each, in the manner of equation~(\ref{eqn:profile-components}). The shape of each component was taken to be Gaussian, with amplitude $a_j$ and width $w_j$:
\begin{equation}
c_j(\phi) = a_j\exp\sbrack{-\frac{(\phi-\mu_j)^2}{2w_j^2}}.
\end{equation}
With this choice of component shape, we have
\begin{align}
	\Gamma_{jk} &= \frac{\sqrt{2\pi}\mkern1mu a_j w_j a_k w_k}{\paren{w_j^2+w_k^2}^{\frac32}}\sbrack{1-\frac{(\mu_j-\mu_k)^2}{w_j^2+w_k^2}}\exp\sbrack{-\frac{\paren{\mu_j-\mu_k}^2}{2\paren{w_j^2+w_k^2}}}\label{eqn:Gamma-gaussian}\\
	\Delta_{jk} &= \frac{\sqrt{2\pi}\mkern1mu a_j w_j a_k w_k}{\paren{w_j^2+w_k^2}^{\frac32}}\paren{\mu_j-\mu_k}\exp\sbrack{-\frac{\paren{\mu_j-\mu_k}^2}{2\paren{w_j^2+w_k^2}}}.\label{eqn:Delta-gaussian}
\end{align}
Amplitudes ($1+b_{ij}$) were drawn from a log-normal distribution with modulation index $m_j$, and phase offsets ($\psi_{ij}$) were drawn from a normal distribution with standard deviation $\bracket[big]{\psi_j^2} = f_jw_j$. The ratio $f_j=\bracket[big]{\psi_j^2}/w_j$, where $w_j$, as here, is the single-pulse width of a particular component, is called the jitter parameter. Below, we will distinguish between different simulated cases by specifying the amplitudes and widths of each component, the separation between components, and the modulation index, $m_j$ and jitter parameter, $f_j$, of each component.

As an initial comparision of simulations with theoretical results, consider the case where pulses are made up of two identical Gaussian components. 
In this case, we have
\begin{align}
	\Gamma_{11} = \Gamma_{22} &= \frac{\sqrt{\pi}a^2}{2w},\\
	\Gamma_{12} = \Gamma_{21} &= \frac{\sqrt{\pi}a^2}{2w}\paren{1-\frac{\delta^2}{2w^2}}\exp\of{-\frac{\delta^2}{4w^2}},\\
	\Delta_{11} = \Delta_{22} &= 0,\\
	\Delta_{12} = -\Delta_{21} &= \frac{\sqrt{\pi}a^2}{2w}\mkern1mu\delta\exp\of{-\frac{\delta^2}{4w^2}},
\end{align}
where $\delta=\mu_1-\mu_2$ is the separation between the components. Equation~(\ref{eqn:tauhat-variance}) then predicts that the variance of the TOA estimates will be given by
\begin{equation}\begin{split}\label{eqn:twocomp-prediction}
	\sigma_\tau^2 &= \frac{(1+m^2)\bracket{\psi^2}(\Gamma_{11}^2+\Gamma_{12}^2)}{2N(\Gamma_{11}+\Gamma_{12})^2} + \frac{m^2\Delta_{12}^2}{2N(\Gamma_{11}+\Gamma_{12})^2}\\
	 &= \curly{\frac{(1+m^2)f^2w^2}{2N}\sbrack{\exp\of{\frac{\delta^2}{2w^2}}+\paren{1-\frac{\delta^2}{2w^2}}^{\!\!2}} + \frac{m^2\delta^2}{2N}}\sbrack{\exp\of{\frac{\delta^2}{4w^2}}+1-\frac{\delta^2}{2w^2}}^{-2}.
\end{split}\end{equation}
Figure~\ref{fig:2comps-sep} compares the predictions of equation~(\ref{eqn:twocomp-prediction}) with simulation results for various values of $\delta$ in three different cases: one with only amplitude variations, another with only phase variations, and a third with both types of variation. With only phase variations, $\sigma_\tau$ is independent of $\delta$, while in cases where phase variations are included, it reaches a maximum when $\delta\approx w$. For reference, a fourth case with neither amplitude nor phase variations, but only a small amount of additive white noise, is shown. In this case, $\sigma_\tau$ is predicted by equation~(\ref{eqn:err-from-weff}). It has some dependence on $\delta$ because $W_{\mr{eff}}$ is a function of the template shape (equation~\ref{eqn:weff}).

A more extensive set of simulation results is shown in Figure~\ref{fig:ncomps-sep}. Shown there are the results of simulations with various numbers of identical components (between 2 and 7), in each of the three primary cases from Figure~\ref{fig:2comps-sep}, along with the predictions of equation~(\ref{eqn:tauhat-variance}). One notable trend visible in Figure~\ref{fig:ncomps-sep} is that, for large separations ($\delta\gg w$), $\sigma_\tau$ decreases as the number of components, $N_c$, is increased. Indeed, it is approximately proportional to $N_c^{-1/2}$. This happens because, for large values of $\delta$, the cross-correlation $\Delta_j$ rapidly approaches $0$, as do all terms in equation~(\ref{eqn:Gamma}) for $\Gamma_j$ except the $j=k$ term, which is independent of $\delta$. Using equation~(\ref{eqn:Gamma-gaussian}), we can see that, for the identical Gaussian components used in the simulations, $\Gamma_j$ approaches $\sqrt{\pi}a^2/(2w)$ for each component. It follows that, in the limit of large $\delta$, we have
\begin{equation}
\sigma_\tau^2 = \frac{(1+m^2)f^2w^2}{NN_c}.
\end{equation}
A similar phenomenon occurs when the components are not identical, complicated only by the fact that the components do not contribute equally to the variance. In general, for sufficiently well-separated components, we have
\begin{equation}\label{eqn:independent-components}
\sigma_\tau^2 = \frac{\sum_{j=1}^{N_c}a_j^4(1+m_j^2)f_j^2}{N\paren[big]{\sum_{j=1}^{N_c}a_j^2w_j^{-1}}^2}.
\end{equation}
In other words, the TOA estimation errors associated with each component combine in a manner weighted by the combination $a_j^2 w_j^{-1}$, and tending to decrease as the number of components increases. However, the trend of decreasing $\sigma_\tau$ with increasing $N_c$ is not universal. As seen in Figure~\ref{fig:ncomps-sep}, in some cases with amplitude variations included, when $\delta\lesssim w$, $\sigma_\tau$ actually increases with increasing $N_c$ at a similar rate. In combination with the fact that $\sigma_\tau$ also depends significantly on the modulation index, $m$, and jitter parameter, $f$, this means that it is generally not possible to predict $\sigma_\tau$ from the number of pulse components without further information.

\bibliography{shape-change-timing}{}

\end{document}